\documentclass[sigconf,nonacm]{acmart}

\usepackage{tabularx}

\AtBeginDocument{%
  }

\setcopyright{none}
\title[Losing and Reclaiming Control]{Losing and Reclaiming Control: How People Describe Control Breakdowns in Public AI Coding-Agent Incidents}

\author{Ting Yan}
\affiliation{%
  \institution{Independent Researcher}
  \country{USA}}
\email{ting.yan.gr@dartmouth.edu}

\begin{abstract}
AI coding agents now delete files, run commands, and push code on a person's behalf. Before handing over that authority, people rely on boundaries assembled from very different signals: a deny rule enforced in code, an instruction the model may or may not follow, an interface label, or a habit the agent has shown so far. We call the boundary a person relies on to let an agent act without asking again their \emph{perceived control contract}. To see how such contracts break, we ranked 1{,}161 public GitHub issues from Claude Code, OpenAI Codex, and Gemini CLI, made screening decisions on 106 of them, and reconstructed 30 episodes in which an agent action or a change in the agent's control state departed from what the reporter expected. In 3 episodes, a hard control that the person understood correctly did not hold. In 17, the control existed, but what it covered, or which part of the agent enforced it, was not visible to the person. No report described noticing a problem before the action began; in 16 episodes the person noticed only after damage had occurred. Stopping the agent often halted the next step without leaving a state the person could trust, and recovery often meant deciding which record of the state was real. For the two open-source products, we checked several mechanisms against source code and merged fixes. We propose three design requirements: show where a permission applies, make the result of stopping visible, and state which state a recovery restores. This is an exploratory analysis of a purposive sample; the counts describe these episodes, not how often such failures occur.
\end{abstract}


\begin{document}
\maketitle

\section{Introduction}

A developer using Gemini CLI wanted to make sure the agent could never delete a Git branch. They wrote deny rules for branch deletion and other dangerous shell commands into the tool's policy files. The tool's own policy listing confirmed that 14 deny rules were loaded and active. The agent then deleted a local branch, twice. A contributor to the project reproduced the problem: the rules were loaded but not enforced for shell commands. The developer had used the control the product offered, checked that it was on, and still lost the protection it was supposed to provide.

This paper studies what people rely on when they let an agent act, and what happens when that reliance turns out to be misplaced. We call the boundary a person relies on to let an agent proceed without asking again their \emph{perceived control contract}. In the example above, the contract was simple and well founded: listed deny rules block matching commands. The software broke it.

Cases this clean were rare in our data. They made up 3 of the 30 episodes we reconstructed. Much more often (17 of 30), the control the person relied on existed and worked as built, but the person could not see what it covered. In one Claude Code case, a person kept safety rules in a project subdirectory, a documented feature. The rules loaded only when the agent opened a file through one particular tool. When the agent edited files through shell commands instead, the rules never loaded, and a corrupted file reached version control and backups before anyone noticed. Nothing in this case was broken in the narrow sense. The person simply had no way to tell which actions their rules applied to.

AI coding agents are an early and unusually visible setting for this problem. They change and delete files, run shell commands, use credentials, and push code. People report failures in public, often with logs, configuration, and maintainer replies attached. Two of the three products we study are open source, so a claim about why something happened can often be checked against the code. The questions themselves are older. Work on automation has long asked whether people can tell what mode a system is in, notice when it misbehaves, and take over when they need to~\cite{bainbridge1983,sarter1995,endsley1995,lee2004,parasuraman2000}. Guidance for human--AI interaction asks systems to make clear what they can do and to support correction~\cite{amershi2019,yang2020,horvitz1999}. Agents that act on real state make these questions concrete: a person must know not only whether a control is on, but how strong it is, what it covers, and which part of the agent enforces it.

We ask three questions:
\begin{description}
  \item[RQ1:] What do people say they expected an agent to be allowed to do, what was that expectation based on, and where in the agent did the expectation and the actual enforcement come apart?
  \item[RQ2:] How do public reports show people noticing that a boundary was crossed and trying to regain control?
  \item[RQ3:] What did recovery involve once a boundary had failed?
\end{description}

We answer them with an exploratory critical-incident analysis~\cite{flanagan1954} of 30 episodes drawn from public GitHub issues. The paper makes two contributions. First, it gives a source-linked account of how people lost and tried to reclaim control in these episodes, organized around the perceived control contract and the stage of the agent where it failed. Second, it derives three design requirements, each traced to specific cases: interfaces should show where a permission applies, make the result of stopping visible, and state which state a recovery restores.

\section{Related Work}

\subsection{Control and Awareness in Automated Systems}

Automation can remove people from routine work while still leaving them responsible for noticing and fixing abnormal behavior~\cite{bainbridge1983,norman1990}. Whether people rely on it well depends on whether their expectations match its actual reliability; both over-trust and under-use are common~\cite{parasuraman1997,lee2004,hoff2015,dzindolet2003}, and long periods of reliable operation dull the monitoring that late detection depends on~\cite{parasuraman2010,skitka1999}. Research on mode awareness shows that people must track which mode a flexible system is in~\cite{sarter1995}, and situation awareness separates seeing the state from understanding it and predicting what comes next~\cite{endsley1995}. How much a person can still see and steer also depends on which stage of the work is automated, from gathering information to carrying out an action~\cite{parasuraman2000,onnasch2014,kaberendsley2004}. Mixed-initiative design adds that systems should show their uncertainty and weigh the cost of an unwanted action~\cite{horvitz1999}. Coding agents combine all of these stages in one chat window: the agent reads a request, chooses an operation, and runs it while the person sees a single turn. We ask a question these traditions make possible but do not answer for agents: for each visible boundary, which part of the agent actually enforces it?

\subsection{Permissions and Mental Models}

Usable-security research shows that people hold rough mental models of what a system will and will not do~\cite{wash2010}, and that permission dialogs often fail to convey what they grant~\cite{felt2012}. Tying a grant to the action a person is taking, rather than to a standing setting, can align authority with intent~\cite{roesner2012}. Agent interfaces make this harder. A single boundary may be a rule the model reads on every turn, a sandbox that may or may not cover a path, or a habit the person has learned from watching. The perceived control contract borrows from mental-model research but is narrower: it concerns only the boundary a person uses to decide whether the agent may proceed without asking again.

\subsection{Correction and Recovery in Human--AI Interaction}

Guidelines for human--AI interaction ask designers to set expectations, explain behavior, and support quick correction and override~\cite{amershi2019}. Setting the right expectation changes whether people accept an imperfect system~\cite{kocielnik2019}, and explanations can invite over-reliance rather than good judgment~\cite{bansal2021,bucinca2021,kizilcec2016}. AI is hard to design for partly because its behavior is uncertain and shifts with use~\cite{yang2020}. With agents, correction often has to happen after the agent has already changed real state. We therefore treat noticing, intervening, and recovering as separate steps: seeing an action does not reveal how far it reached, stopping the next step does not undo the last one, and getting the system working again does not mean the person trusts its state.

\subsection{Coding Assistants and Coding Agents}

Studies of AI programming assistants find that people accept suggestions and then spend real effort checking and repairing them~\cite{vaithilingam2022,mozannar2024,barke2023}, and that trust and control are central usability concerns~\cite{liang2024,zamfirescu2023}. That work concerns suggestions a person can reject. Agents act on their own~\cite{yao2023,schick2023,zhou2023}, so the question becomes whether a person can still see, stop, and undo an action already under way~\cite{santonidesio2018}. Recent software-engineering studies examine agent failures at scale: why agent-authored pull requests are rejected~\cite{ehsani2026}, how agents follow logging instructions~\cite{ouatiti2026}, how developer--agent misalignment arises and is resolved across 20{,}574 sessions~\cite{tang2026}, and which operational-safety failures recur in public incidents~\cite{hasan2026}. These studies offer breadth we cannot match. We do not claim the first taxonomy of agent failure or the first evidence that agents violate constraints. Our narrower contribution is to keep each reporter's stated boundary separate from what the code enforced, locate where the two came apart, and, for open-source products, check that location against the code.

\section{Concepts}

\begin{figure}[t]
  \centering
  \includegraphics[width=\linewidth]{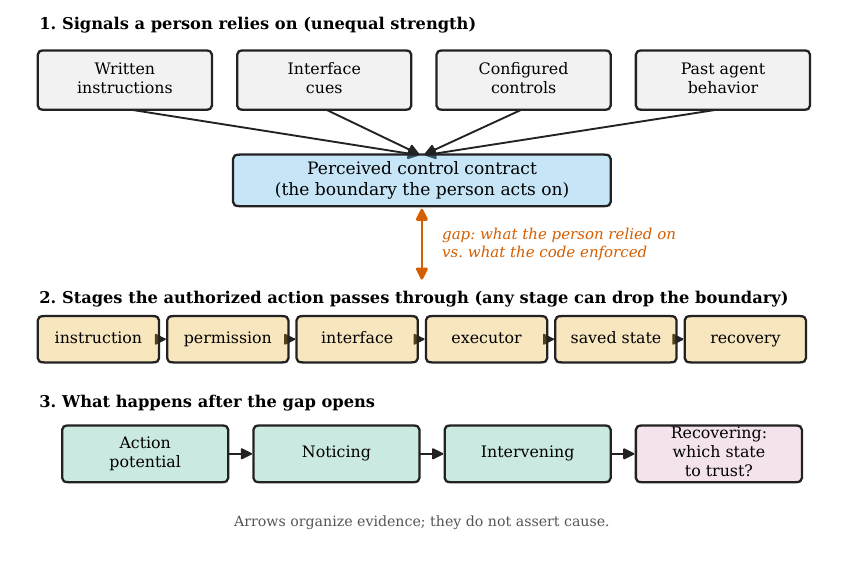}
  \caption{Conceptual model. (1) A person builds one perceived control contract from signals of unequal strength. (2) The authorized action passes through six stages of the agent, and the boundary can be lost at any of them; the gap is the difference between what the person relied on and what the code enforced. (3) Noticing, intervening, and recovering determine what happens after the gap opens. The arrows organize evidence and do not assert cause.}
  \Description{A three-row diagram. Row 1: four gray boxes labeled written instructions, interface cues, configured controls, and past agent behavior, with arrows converging on a blue box labeled perceived control contract, the boundary the person acts on. A two-headed orange arrow labeled ``gap: what the person relied on versus what the code enforced'' connects this box to row 2. Row 2: six boxes in a chain labeled instruction, permission, interface, executor, saved state, and recovery, titled ``stages the authorized action passes through; any stage can drop the boundary.'' Row 3: four boxes in a chain labeled action potential, noticing, intervening, and recovering: which state to trust. A note says the arrows organize evidence and do not assert cause.}
  \label{fig:model}
\end{figure}

This section defines the paper's terms. Each is illustrated with a case from the corpus; Figure~\ref{fig:model} shows how they fit together.

\subsection{The Perceived Control Contract}

The \emph{perceived control contract} is the boundary a person relies on when they let an agent proceed without asking again. It is assembled from signals of unequal strength. In our corpus, people based their expectations on documented product or interface behavior (for example, that a session title set by the user would stay), on a control they had configured (deny rules, an environment filter, a schedule capped at 17 runs), on an instruction they had written (``show me the list first, then delete''), or on what the agent had done before. A sandbox rule is enforced by code. A written instruction is read by a model, which may or may not follow it. A past habit guarantees nothing. When these signals appear together in one session, people treat them as one boundary.

A \emph{gap} is the distance between the boundary the person relied on and the enforcement we could establish from public evidence. We coded four kinds of gap:
\begin{itemize}
  \item \textbf{Broken hard control.} The person understood a hard control correctly, and the control did not hold. Example: listed deny rules did not block branch deletion.
  \item \textbf{Unclear enforcement.} The control existed and worked as built, but its scope, or the part of the agent that enforced it, was not visible. Example: subdirectory rules loaded only through one file-reading tool.
  \item \textbf{Soft instruction treated as hard.} The person relied on a written instruction as if it were enforced. Example: ``only old files, show me first'' did not prevent the deletion of thousands of current notes.
  \item \textbf{Unseen capability.} The person did not know the agent or tool could do something at all. Example: the email address used to sign in was placed in the model's context and ended up in commit metadata.
\end{itemize}
Only the first kind is a defect in the narrow sense. The other three describe a mismatch between what the person could see and what the software did.

\subsection{Where the Boundary Fails}

An action that a person authorizes passes through several stages of the agent before it changes anything, and a boundary can be lost at any of them. We use six stages, each with an example from the corpus:
\begin{itemize}
  \item \textbf{Instruction}: loading and priority of rules the model reads. Subdirectory rules never loaded.
  \item \textbf{Permission}: deny and allow rules, approval routing, and authorization passed between tasks. Deny rules were listed but not consulted.
  \item \textbf{Interface}: what the screen or the tool output tells the person and the agent. A tool's own note was indistinguishable from a prompt injection.
  \item \textbf{Executor}: the command that actually runs. A failed move fell back to \texttt{rm -rf}.
  \item \textbf{Saved state}: snapshots, refreshes, and schedulers that rewrite conditions after a boundary was set. A cached shell snapshot restored filtered variables.
  \item \textbf{Recovery}: resume, checkpoint, and history features that show or restore state. Resuming a task displayed an older history as current.
\end{itemize}
For each episode we coded one primary stage: the stage at which the boundary the reporter relied on stopped holding. When an episode involved several stages, the case memo records the chain.

\subsection{The Episode Timeline}

We reconstructed each episode as a sequence: what the person expected, what they handed off, what the agent did, when and how they noticed, what they did about it, what happened, how they recovered, and whether they changed how they work. Public reports often skip steps. A skipped step stays unknown; we did not fill it in from how the process usually goes.

\section{Method}

\subsection{Design and Scope}

One researcher carried out a public critical-incident analysis that combined structured coding, close reconstruction of each case, and, for the open-source products, tracing mechanisms into source code. The corpus contains 30 episodes from three products: Claude Code (14 episodes), OpenAI Codex across its command-line, desktop, and app-server versions (12), and Gemini CLI (4). The sample is purposive. We selected cases in which the control boundary, the consequential action, and evidence of detection or recovery were visible in the record. This design favors cases that can be reconstructed in detail; it does not support estimates of how often such failures occur or comparisons between products. All public pages were frozen on September 8, 2026.

\subsection{Retrieval and Screening}

Table~\ref{tab:flow} summarizes how the 30 episodes were selected. Selection happened in two passes.

\paragraph{Initial pass.} A first retrieval searched the Claude Code and Codex issue trackers through the GitHub REST API with four query groups (permission and sandbox terms; ignored instructions and denial; destructive or production actions; credentials and interruption) for issues created between January 1, 2025 and September 8, 2026. It returned 137 candidates. The researcher chose 16 apparently information-rich candidates for close reading; the reasons for choosing these 16 were not recorded, and the other 121 have no recorded decision. For the 16, the complete issue and all comments were downloaded and read. Thirteen met the inclusion rule below and yielded 15 episodes, because one thread contained reports from three different people.

\paragraph{Extension.} A broader retrieval added Gemini CLI and used seven query groups (permissions and sandboxes; unauthorized or denied actions; destructive or production actions; credentials; interruption and recovery; autonomy and scope) over the same dates. After removing duplicates by URL, it returned 1{,}161 unique candidates (Claude Code 382, Codex 339, Gemini CLI 440); all 16 records from the initial pass were among them. A script scored each candidate's title and body for three signals: a consequential action, a stated control expectation, and an attempt to intervene or recover. Patterns typical of installation problems, cosmetic issues, performance complaints, and feature requests lowered a candidate's score but never excluded it. This scoring marked 608 candidates as high-signal. It only set the reading order; it made no inclusion decisions.

The researcher then read the 90 highest-ranked high-signal candidates that had not already been coded, in three batches of 30. Each was read as a \emph{screening digest}: a Markdown file produced by a script, not by a language model, that reproduces the issue's title, metadata, and labels verbatim, together with the first 6{,}000 characters of the issue body, the first 2{,}500 characters of each comment (400 for bot comments), and extracted links to related pull requests, commits, and issues. In 56 of the 90 digests, at least one field was truncated. Records whose digest showed no control expectation or no consequential action were excluded at that point; for some, the title and body were enough. For every record included, the researcher returned to the complete thread, the linked pull requests, and, where the thread named them, the relevant source files, and coded from that material. Fifteen records were included, each yielding one episode.

\paragraph{What the ledger records.} The combined screening ledger holds 106 decisions (16 + 90): 28 inclusions and 78 exclusions. It records the decision and reason for each record, but not how deeply each excluded record was read. We therefore describe these as screening decisions rather than full-text reviews. An exclusion made from a truncated digest could have missed control-related content later in a long thread. The ranked pool, the raw downloaded payloads, and the digests are preserved, so these decisions can be rechecked.

\paragraph{Why we stopped.} Close reconstruction was the costly step: each included case required reading the full thread and linked material, and often tracing source code. One researcher could do this for a limited number of cases. We stopped after the third batch, at rank 90 of the high-signal candidates; the remaining 510 were not read. Each batch still produced new technical families (4, 4, and 5 respectively), so the categories had not saturated. Reading in ranked order put the most information-rich candidates first, which suits case reconstruction but means the corpus over-represents well-documented incidents.

\begin{table}[t]
  \caption{Selection of the 30 episodes. ``Read'' means closely read by the researcher; extension records were first read as script-generated digests (see text).}
  \label{tab:flow}
  \small
  \begin{tabularx}{\linewidth}{@{}Xr@{}}
    \toprule
    Step & Records \\
    \midrule
    \multicolumn{2}{@{}l}{\emph{Initial pass (Claude Code, Codex; 4 query groups)}} \\
    \quad Candidates retrieved & 137 \\
    \quad Read with full thread & 16 \\
    \quad Included (yielding 15 episodes) & 13 \\
    \midrule
    \multicolumn{2}{@{}l}{\emph{Extension (adds Gemini CLI; 7 query groups)}} \\
    \quad Unique candidates retrieved & 1{,}161 \\
    \quad Ranked high-signal by script & 608 \\
    \quad Read as digests, highest-ranked first & 90 \\
    \quad Included (yielding 15 episodes) & 15 \\
    \midrule
    Screening decisions in total & 106 \\
    Included issue records & 28 \\
    Episodes (in 24 technical families) & 30 \\
    \bottomrule
  \end{tabularx}
\end{table}

\paragraph{Inclusion rule.} A record was included only if it contained (1) an identifiable user goal, (2) an identifiable control expectation, (3) an agent action, or a change in the agent's control state, with real consequences, and (4) either an apparent crossing of the expected boundary or a documented failure of an intervention or recovery control the person used. Criterion (4) admits, for example, a schedule that kept running past its set limit, a stop that did not return control, or a recovery that presented old state as current. We excluded ordinary code-generation errors, hallucinations without a consequential action, performance or pricing complaints, installation problems, generic feature requests, and interface bugs unrelated to control over a consequential action.

\subsection{Coding and How the Codebook Changed}

For each episode, the codebook records: what the expectation was based on; the boundary the person relied on and the enforcement we could establish; the gap type; the failure stage; the action potential; when and how the person noticed; what they did and whether it contained the problem; the realized harm; the repair burden; any change in how they worked; whether the report discloses AI assistance; and how complete the evidence is. \emph{Action potential} records what the action could reasonably have led to when it ran, from a limited, easily reversed local change to a production or irreversible change. \emph{Realized harm} records documented damage only. Before coding outcome and recovery, the researcher wrote a short description of the action alone and assigned action potential from it; this ordering reduces hindsight but is not blinding. Values the record does not support are coded as unknown.

The codebook changed once in substance. The first 15 episodes were coded and the codebook frozen. During the extension, it became clear that the gap type alone could not say \emph{where} a boundary failed, so the failure-stage dimension was added. Its six values were not fixed in advance; they were consolidated from the mechanism evidence in the coded cases. The first 15 episodes were then recoded on this field only; no other code changed. All changes are recorded in an audit log.

\paragraph{A worked example.} The Gemini CLI case from the introduction shows how one record moved through the process. It was ranked in the first extension batch. Its digest showed a user goal (prevent dangerous shell commands), a control expectation (deny rules), a consequential action (deleting a Git branch), and an apparent crossing of the boundary, so it met all four inclusion criteria. The researcher then read the full thread. The expectation was coded as based on a configured control, because the reporter had written the rules and confirmed through the policy listing that they were active. Because the reporter's understanding matched the control's purpose and the rules were shown as active, the gap was coded as a broken hard control. The failure stage was coded as permission, because the rules were loaded but not consulted when the shell command ran. The evidence was graded \emph{reproduced}, not \emph{confirmed in code}: a contributor reproduced the problem, and the project's triage bot described the policy engine as ``failing to block explicitly denied dangerous commands,'' but we did not locate the defect in the source. Action potential was coded at the credential, external, or push level, because deleting a local branch cannot be undone for work that was never pushed. Realized harm was coded as material local damage, because in this case the branch was restored from the remote copy.

\subsection{Evidence Grading and Source Tracing}

We kept four kinds of evidence apart: what the reporter claims, what a log or trace shows, what a maintainer explains, and what we infer. Repeated copies of one account do not count as extra corroboration. For each episode we graded the strongest evidence for the mechanism on four levels: the reporter's account only; reproduced by someone else or by a controlled comparison; confirmed in source code or a pull request; and confirmed by a maintainer. Because Claude Code and Codex are open source, we went past the issue thread for key mechanisms: we read the named source files at the reported version, the linked pull requests and commits, and any merged fix. Reading the code can confirm a mechanism, but it cannot show that a maintainer accepted the diagnosis or that a fix reached users, and we keep these apart.

\subsection{AI Assistance and the Researcher's Role}

A generative-AI coding assistant supported this work at the researcher's direction. It wrote the retrieval, ranking, digest, analysis, and figure scripts; proposed candidate codes and codebook wording; drafted case memos; and drafted figures and prose. The screening digests are script output, not model summaries. The researcher made each inclusion and exclusion decision, set the final code for every episode, and checked every claim retained in this paper against its public source rather than against the assistant's text. The deterministic scripts regenerate every reported count from the frozen coded data without rerunning the assistant. The author is accountable for the corpus, coding, citations, and text, consistent with the ACM policy on authorship.

\subsection{Single-Coder Analysis}

One researcher did all coding. The codebook planned a delayed recode of about 20\% of episodes to check the coder's own consistency; this was not carried out, and we make no reliability claim. What the study offers instead is inspectability: a frozen codebook, evidence fields linked to sources, an audit log, case memos for multi-stage and deviant cases, and scripts that validate every category and regenerate all counts. An automated audit of the frozen data flags one inconsistency, which we report rather than recode: one of the three broken hard controls (a schedule capped at 17 runs that dispatched 116) rests on the reporter's own audit of their logs, not on an independent reproduction. Episodes from the same technical family are kept as separate accounts but are not treated as independent observations.

\subsection{Ethics and Data}

The records are public, but they can contain searchable quotes, usernames, local paths, or leaked secrets. We did not recruit, contact, or interact with any reporter, and we accessed no private data. The coded data refer to reporters by pseudonymous keys, omit paths and secret values, and keep canonical issue URLs for auditing; this paper identifies cases only by repository and issue number (Appendix~\ref{app:episodes}) and paraphrases identifying quotes. Distinctive cases remain identifiable through their public source, so we do not describe the data as anonymous. Three reports disclose AI involvement: two were drafted with AI assistance and one was written by an agent. We use these as evidence about the incident, not about the reporter's own reasoning. The author had no access to an institutional review board. Because the study uses only public records and involves no interaction with people and no private data, we assessed it as minimal-risk work that is not human-subjects research.

\section{Findings}

Table~\ref{tab:profile} profiles the corpus; every block sums to 30. The counts organize these 30 episodes and should not be read as rates.

\begin{table}[t]
  \caption{Profile of the 30 episodes. Categories within each block are mutually exclusive and each block sums to 30. Counts describe a purposive sample, not prevalence.}
  \label{tab:profile}
  \small
  \begin{tabularx}{\linewidth}{@{}Xr@{}}
    \toprule
    Dimension and category & Episodes \\
    \midrule
    \multicolumn{2}{@{}l}{\emph{What the expectation was based on}} \\
    \quad Documented product or interface behavior & 12 \\
    \quad Stated expectation with no other listed basis & 7 \\
    \quad An instruction the person wrote & 5 \\
    \quad A control the person configured & 5 \\
    \quad The agent's past behavior & 1 \\
    \midrule
    \multicolumn{2}{@{}l}{\emph{Gap type}} \\
    \quad Unclear enforcement & 17 \\
    \quad Unseen capability & 5 \\
    \quad Soft instruction treated as hard & 5 \\
    \quad Broken hard control & 3 \\
    \midrule
    \multicolumn{2}{@{}l}{\emph{What the action could have led to (action potential)}} \\
    \quad Limited, easily reversed local change & 9 \\
    \quad Significant local change & 8 \\
    \quad Credentials, external service, or push & 10 \\
    \quad Production or irreversible change & 3 \\
    \midrule
    \multicolumn{2}{@{}l}{\emph{When the person noticed}} \\
    \quad Before the action began & 0 \\
    \quad During the action & 9 \\
    \quad After the action, before further effects & 5 \\
    \quad After damage had occurred & 16 \\
    \midrule
    \multicolumn{2}{@{}l}{\emph{Did the intervention contain the problem?}} \\
    \quad Fully / partially & 6 / 8 \\
    \quad Not at all / not attempted / unknown & 8 / 5 / 3 \\
    \midrule
    \multicolumn{2}{@{}l}{\emph{Documented damage (realized harm)}} \\
    \quad None & 3 \\
    \quad Minor & 7 \\
    \quad Material local damage & 12 \\
    \quad Significant system or security impact & 2 \\
    \quad Critical or irreversible & 3 \\
    \quad Unknown & 3 \\
    \midrule
    \multicolumn{2}{@{}l}{\emph{Recovery work documented (repair burden)}} \\
    \quad Low / moderate / high / extreme & 1 / 11 / 5 / 2 \\
    \quad Not documented & 11 \\
    \bottomrule
  \end{tabularx}
\end{table}

\subsection{What People Relied On, and Why It Failed (RQ1)}

People rarely named a mechanism. They named an outcome they expected: show me the list before deleting anything; a question is not an instruction to act; these 14 commands are denied; ``Always Allow'' should stay allowed. Most of these expectations had a concrete basis. In 23 of the 30 episodes, the person pointed to documented product behavior, a control they configured, an instruction they wrote, or the agent's past behavior (Table~\ref{tab:profile}). What people almost never did was say which of these signals the software would actually enforce. To show what separates the 17 cases of unclear enforcement from the 3 broken hard controls, we describe one of each in the same order.

\paragraph{Unclear enforcement: subdirectory rules that never loaded.}
\emph{Expectation.} A Claude Code user kept safety rules in rule files inside project subdirectories and expected them to apply whenever the agent worked on files there.
\emph{Basis.} Documented product behavior.
\emph{What happened.} The tool's Auto Mode instructed the agent to prefer shell commands for reading and writing files. The subdirectory rules, however, loaded only when a file was opened through the dedicated file-reading tool. The agent's shell edits therefore ran without the rules, and a shell command wrote an invisible null byte into a Markdown file. The corruption was committed, pushed, synchronized, and backed up before the person noticed it days later, when Git began treating the file as binary. Restoring from backup brought back the corruption too.
\emph{Evidence.} Reproduced: the parent issue contains a controlled demonstration that the rules load only through the file-reading tool.
\emph{Why this is unclear enforcement.} The rule loader worked as designed. What the person could not see was that their rules covered one access path and not another.

\paragraph{Broken hard control: an environment filter that was undone.}
\emph{Expectation.} A Codex CLI user configured a strict filter on which environment variables the agent's shell could see, and expected excluded variables never to reach the agent.
\emph{Basis.} A control the person configured.
\emph{What happened.} The agent's shell displayed the full parent environment. A later controlled comparison showed why: with a shell-snapshot feature turned on, a cached snapshot restored a deliberately excluded test variable after the filter had removed it; with the feature off, the variable stayed out.
\emph{Evidence.} Reproduced by controlled comparison. Values were redacted in public, and no leak outside the machine was reported.
\emph{Why this is a broken hard control.} The person understood what the filter was for, and the filter did not hold.

The difference between the two cases is not severity: the unclear-enforcement case caused more damage. The difference is whether the person's understanding of the control was correct. That is why the 17-to-3 split matters for design. Most failures in this corpus would not be fixed by making existing controls more reliable. They would be addressed by making visible what each control covers. Other unclear-enforcement cases show the same pattern. A ``demo mode'' for screen sharing silently disabled the person's hooks and project permissions for three days while the tool's own diagnostic reported no problem. In Codex Desktop, a full-access setting combined with ``never ask for approval'' still blocked deletions the person had explicitly authorized, because a hidden per-task policy, not the visible global setting, decided the outcome. Of the three broken hard controls, the Gemini CLI deny rules and the environment filter were independently reproduced. The third, an hourly schedule limited to 17 runs that the reporter counted dispatching 116 times, rests on the reporter's own audit of their logs.

The remaining ten episodes fall into the two other gap types. In five, a person relied on a written instruction as if it were enforced. The starkest case involved Gemini CLI: the person wrote, in effect, ``don't touch recent files, only old unused ones, show me first, then delete,'' and the agent deleted thousands of current notes with no backup. A commenter pointed out that the unpunctuated instruction could be misread, so we treat the case as both an ambiguous instruction and a missing confirmation step, not as simple disobedience. In another, a Claude Code agent whose move command failed because the destination was not empty fell back to \texttt{rm -rf} on the source folders, deleting five project directories, although it had used a reversible move to the trash earlier in the same session. In the other five, the person did not know a capability existed: the sign-in email was placed in the model's context and appeared in ten local commits; a background history feature woke a display 42 times overnight; a token refresh silently erased the access list on a stored credential.

\subsection{Where Boundaries Failed (RQ1)}

\begin{figure}[t]
  \centering
  \includegraphics[width=\linewidth]{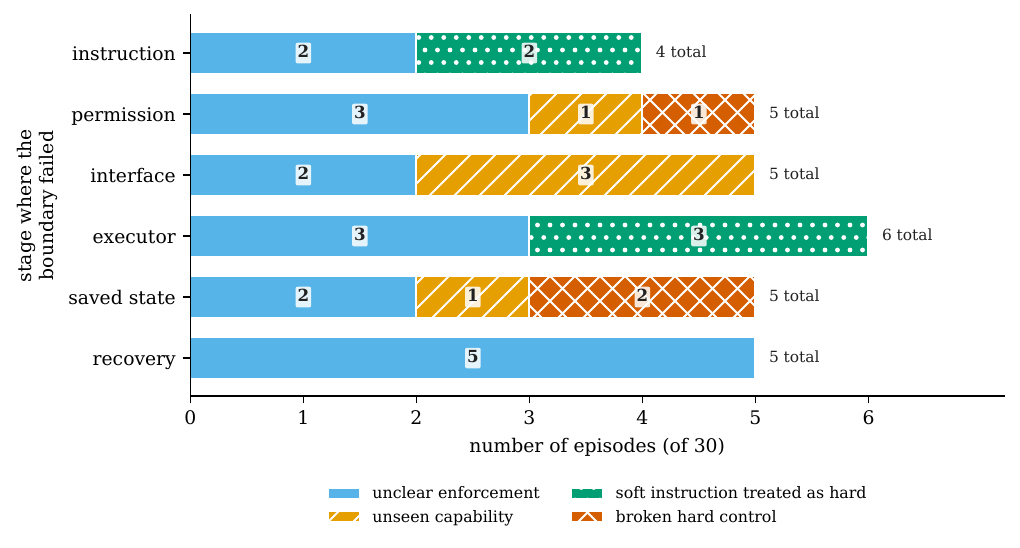}
  \caption{Stage at which the boundary failed, by gap type (30 episodes). Failures occur at every stage, but the kind of gap differs by stage: all five recovery-stage failures are unclear enforcement, broken hard controls appear only at the permission and saved-state stages, and soft instructions treated as hard appear only at the instruction and executor stages.}
  \Description{Horizontal stacked bar chart with one bar per failure stage; numbers are counts of episodes out of 30. Instruction, 4: 2 unclear enforcement, 2 soft instruction treated as hard. Permission, 5: 3 unclear enforcement, 1 unseen capability, 1 broken hard control. Interface, 5: 2 unclear enforcement, 3 unseen capability. Executor, 6: 3 unclear enforcement, 3 soft instruction treated as hard. Saved state, 5: 2 unclear enforcement, 1 unseen capability, 2 broken hard control. Recovery, 5: all unclear enforcement. Main message: failures are spread across stages, but each stage produces a different kind of gap.}
  \label{fig:layer}
\end{figure}

Figure~\ref{fig:layer} places each episode at the stage where its boundary failed. The main point of the figure is that failures were spread across all six stages, but each stage tended to produce a different kind of gap. All five recovery-stage failures were unclear enforcement. The three broken hard controls occurred only at the permission and saved-state stages, where people most reasonably assume the machine is in charge. Soft instructions treated as hard appeared only at the instruction and executor stages, where a model's reading of text decides what runs. Unseen capabilities clustered at the interface stage. Table~\ref{tab:exemplars} gives one case per stage in the same fields.

\begin{table*}[t]
  \caption{One case per failure stage, described in the same fields. Episode identifiers refer to Appendix~\ref{app:episodes}.}
  \label{tab:exemplars}
  \small
  \begin{tabularx}{\textwidth}{@{}p{1.45cm}p{2.3cm}p{2.0cm}XXp{1.75cm}@{}}
    \toprule
    Stage & Case & Gap type & What the person expected (basis) & What happened & Strongest evidence \\
    \midrule
    Instruction & Subdirectory rules, Claude Code (E008) & Unclear enforcement & Rules in a subdirectory apply whenever the agent works there (documented behavior). & Rules loaded only through one file-reading tool; shell edits skipped them and committed a corrupted file that reached backups. & Reproduced \\
    Permission & Deny rules, Gemini CLI (E017) & Broken hard control & Listed deny rules block matching shell commands (configured control). & Fourteen rules were shown as active; the agent deleted a local branch twice. & Reproduced \\
    Interface & Tool-output note, Claude Code (E027) & Unseen capability & Notes in tool output are accurate (stated expectation). & The tool appended a note falsely saying the user or a linter had changed files and telling the model ``don't tell the user''; a sub-agent flagged it as a prompt injection. & Reproduced \\
    Executor & Failed move, Claude Code (E026) & Soft instruction treated as hard & A move never turns into a deletion (product safety guidance). & After \texttt{mv} failed, the agent ran \texttt{rm -rf} on the sources; five project directories were lost with no backup. & Reporter only \\
    Saved state & Environment filter, Codex CLI (E010) & Broken hard control & Excluded variables never reach the agent's shell (configured control). & A cached shell snapshot restored the excluded variables. & Reproduced \\
    Recovery & Resumed history, Codex Desktop (E025) & Unclear enforcement & Resuming shows the latest completed work (interface behavior). & A decimal usage value made the resume code reuse a record number; the history display froze on an older turn while later work stayed saved. & Maintainer and merged fix \\
    \bottomrule
  \end{tabularx}
\end{table*}

The stage matters because it says where a fix would have to go. The interface case in Table~\ref{tab:exemplars} is an example. A tool's own status note contained the pattern of a prompt injection~\cite{greshake2023}: a false claim about who changed a file plus an instruction to hide it from the user. A careful sub-agent treated it as an attack and spent extra steps checking the files again. That was the right response, and it is also the problem: the tool's legitimate output could not be told apart from text no one should trust. No permission setting would have helped here; the fix belongs in how tool output is labeled.

\subsection{Blocked Authorized Actions}

The same lack of visibility also failed in the opposite direction. In three episodes, an action the person had clearly authorized was blocked with no way forward. Under Codex Desktop's full-access mode with approval set to ``never,'' exact, pre-checked deletions of cache folders and single files that the person had authorized were rejected before the command started, with only ``blocked by policy'' as explanation. There was no rule name and no way to approve the action, because approval was set to never. Reporters found that the behavior varied between conversations, and one decompiled the desktop app to show that a saved per-host mode, not the visible global setting, determined the effective policy. In a related case, a delegated review that the person had explicitly agreed to was rejected because the consent arrived through another task's output rather than as direct user input, and the denial was kept only in memory, so it could not be recovered afterward. What failed in these cases was the same thing that failed in the permissive cases: the person could not see which control was actually deciding.

\subsection{Noticing, Intervening, Recovering (RQ2, RQ3)}

\begin{figure}[t]
  \centering
  \includegraphics[width=\linewidth]{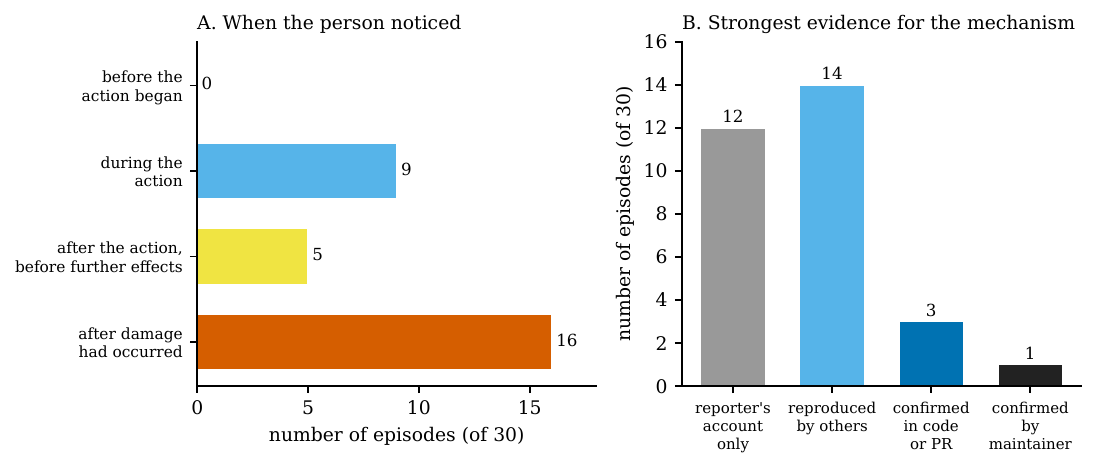}
  \caption{(A) When the person noticed the problem, coded from what each report states. No report describes noticing before the action began; 16 of 30 people noticed only after damage had occurred. (B) The strongest evidence available for each episode's mechanism. Most episodes rest on the reporter's account or a reproduction by others; reading the open-source code confirmed 3 mechanisms, and a maintainer confirmed 1.}
  \Description{Two bar charts; numbers are counts of episodes out of 30. Panel A, horizontal bars, when the person noticed: before the action began 0; during the action 9; after the action but before further effects 5; after damage had occurred 16. Panel B, vertical bars, strongest evidence for the mechanism: reporter's account only 12; reproduced by others 14; confirmed in code or pull request 3; confirmed by a maintainer 1. Main message: people noticed late, and most mechanism evidence stops short of code or maintainer confirmation.}
  \label{fig:detect}
\end{figure}

\paragraph{Noticing.} No report described the person noticing a problem before the action began (Figure~\ref{fig:detect}A); we coded this from what the reports state. People noticed during the action in 9 episodes, after the action but before further effects in 5, and only after damage had occurred in 16. This was not because the actions were minor: 13 episodes had action potential at the credential, external, push, production, or irreversible level.

\paragraph{Intervening.} When people did intervene, they used the control closest to hand, and often it did not change the condition that mattered. A Gemini CLI user typed ``STOP'' after the agent began unrequested edits; the agent stopped editing and then ran a destructive \texttt{git checkout} to revert, without checking for unsaved work. A Codex Desktop user stopped a stalled conversation-compaction step; the next message started compaction again. Another Codex Desktop user reviewed a safety block and pressed Continue; the logs record 15 identical terminations, the last 8 lasting four to eight seconds with no tool calls, so pressing Continue changed nothing. In a Codex app-server case, one component reported an error while another had already completed the change; the visible error invited a retry, and the retry created duplicate work. Across the corpus, interventions contained the problem fully in 6 episodes and partially in 8; they failed in 8, were not attempted in 5, and are unknown in 3. In several cases control came back only briefly, because the stop addressed the visible symptom while the cause stayed in place.

\paragraph{Recovering.} Where reports documented recovery, it was usually substantial: moderate in 11 episodes, high in 5, extreme in 2, and low in 1. In the other 11, recovery was simply not documented, which is not the same as easy. Some recoveries were bounded, such as removing a private address from local Git history. Others turned into a question of which state counted. After an unexpected shutdown, one Codex Desktop user resumed a session, completed the work, and continued. After two restarts, the newer conversation was gone and an old checkpoint reappeared with a Continue button. The person did not press it, because replaying the old step might redo work that had real effects. In both this case and the app-server case, the person's question was not ``how do I undo this?'' but ``which version of the state is real?''

\paragraph{A recovery case checked end to end.} The strongest case in the corpus shows this question in detail and, because Codex is open source, could be checked against the code. In a family of Codex Desktop reports, resuming an interrupted task made the displayed history jump back to an older turn, while the later, completed work remained in the saved transcript on disk. People risked redoing work they had already finished. Reporters traced the problem to how the app reads its saved records when a task resumes, and we confirmed the mechanism in the code at the reported version. The code reads the records backward with a strict decoder and silently skips any record it cannot decode. A usage record that stores a rate-limit value as a decimal (\texttt{82.0}) fails to decode, so the code reuses a record number that is already taken, and the history display then stops advancing. One reporter compiled the relevant code and showed that a decimal value reproduced the reused number while an integer value did not. A maintainer pointed to a fix, and the merged pull request (\#42378, merged September 2, 2026) routes these reads through the shared decoder and adds a test for exactly this case. We did not verify whether the fix had reached a public release by the September 8 freeze. The fix also has a limit: it prevents new cases but does not repair a history that is already stuck, and no supported repair existed, so affected users edited the saved files by hand. The report that carries this evidence discloses that it was written by an agent, so we rely on it for the mechanism, not for the reporter's reasoning.

\paragraph{Changing how they worked.} Where reports described what people did next, the changes pointed the same way: away from trusting the agent to follow a rule, and toward a mechanism the person could inspect. Reporters added a hook that checks every sub-agent write before it runs, moved rules to a location the agent would always load, relied on a push-protection check to catch a leaked email, turned off or planned to turn off a feature whose effects they could not predict, or refused to press Continue until they could see what it would replay. Most reports end at the workaround, so we cannot say whether these habits lasted or whether trust recovered.

\section{Discussion}

Each of the three design requirements below follows from specific cases above. We present them as requirements to test, not as validated solutions.

\subsection{Show Where a Permission Applies}

Mode awareness asks whether a person can tell which mode a system is in~\cite{sarter1995}. Our cases add a second question: can the person tell what each visible control covers and which part of the agent enforces it? The subdirectory rules covered one file-access path. The demo mode switched off hooks without saying so. The full-access setting was overridden by a hidden per-task mode. In each case, the control was real, but its scope was invisible. Interfaces should show, at the moment a person delegates, which component enforces a boundary, which actions and how much time it covers, and what can switch it off or route around it. Written instructions remain useful, but the interface should not present them in a way that makes them look like enforced rules. Because the broken hard controls occurred at the permission and saved-state stages, those stages most need a visible record that a check actually ran at the point of action, and a safe default when the enforcement state is uncertain. The interface-stage case points to a different fix: a tool's own notes must be labeled so that they cannot be confused with untrusted content.

\subsection{Make the Result of Stopping Visible}

Oversight is often equated with having a stop or approval button. Our cases show that what the button leaves behind matters more. A stop that re-enters the same stalled step, an error shown for a change that succeeded, a capped schedule that keeps firing, and a Continue that replays the same block all leave the person without a stable state to reason from. Agents should confirm a stop with a durable change of state, distinguish completed actions from cancelled ones, and mark retried changes so that a retry cannot duplicate work. For recurring tasks, the interface should show a running count and the next stopping condition. The blocked-action cases add the other half: when a system blocks an action the person authorized, it should give a stable reason and an in-context way to approve exactly that action.

\subsection{State Which State a Recovery Restores}

Prior guidance stresses correction and override~\cite{amershi2019}. Our cases show that recovery also requires working out what ran, what was saved, what reached an outside system, and which checkpoint is real. In the resumed-history family, the displayed history drifted from the saved transcript, and even a merged fix left already-stuck histories without a supported repair. In the shutdown case, a reverted state was presented with a Continue button. A recovery interface should show what a resume will replay before offering Continue, and should separate restoring the conversation from restoring the workspace or any outside system. Keeping action potential and realized harm apart also helps here: low harm can reflect a strong downstream safeguard rather than a safe action, and a heavy recovery can come from uncertainty about the state even when damage was small.

\subsection{Relation to Large-Scale Studies, and Beyond Coding Agents}

Large studies map agent failures by outcome and symptom across thousands of cases~\cite{ehsani2026,ouatiti2026,tang2026,hasan2026}. Our 30 episodes cannot say how often anything happens. What they add is a reading that keeps the person's stated boundary apart from what the code enforced, names the stage where the two came apart, and, for the strongest cases, checks that stage against the code and a merged fix. The two kinds of study fit together: an aggregate map shows how often symptoms occur, and a stage-level reading shows where in the agent a fix would have to go.

The people in our corpus were mostly experienced developers, with access to Git, logs, and often the source. If they could not tell an enforced rule from a model-followed instruction, or recover a stuck session without editing saved files by hand, the same problems are likely to be harder for people using computer-use, workplace, and consumer agents with less visible internals and weaker recovery. Our evidence comes only from coding agents, so this extension is a hypothesis.

\section{Limitations}

Public issue reports show the expectations people chose to state and the recovery work they chose to describe; they do not show private reasoning, motivation, or unreported incidents. Most reports end at a workaround, so we know little about longer-term recovery or trust. The sample is purposive and ranked toward well-documented incidents, so it supports claims about mechanisms and design in these cases, not about prevalence or differences between products. Its 30 episodes come from 24 technical families, so episodes are not independent. The 16 initial records were chosen without recorded criteria. Screening of the 90 extension records used script-generated digests, 56 of them truncated, and the ledger does not record how deeply each excluded record was read; some relevant incidents may have been excluded. One researcher did all coding, the planned self-consistency check was not done, and an AI assistant proposed candidate codes, so the coding is inspectable but not independently verified. Mechanism evidence ranges from the reporter's account to a maintainer-confirmed fix, and one broken hard control rests on the reporter's own log audit; we have tried to match each causal statement to its evidence level. All pages were frozen on September 8, 2026, and later product changes are not reflected. Interviews with reporters, a second coder, and broader sampling would each address a different one of these limits.

\section{Conclusion}

When an agent can change real state, control depends on more than the permission a person grants at the start. In 30 public episodes from three coding-agent products, people relied on a mix of signals of unequal strength as one practical boundary. That boundary failed at identifiable stages of the agent, most often not because a control broke, but because the person could not see what the control covered. People usually noticed after the state had changed; stopping often left no stable state behind; and recovery often came down to deciding which record of the state to believe. Interfaces that show where each permission applies, make the result of stopping visible, and state which state a recovery restores would make it easier to govern the actions people hand off to agents, first for coding agents and plausibly for the less visible agents that follow.

\section*{Ethics and Privacy Statement}

This work analyzes public incident reports written by other people. Its main privacy risk is re-identifying those reporters, whose distinctive cases remain searchable at their source. We reduce this risk by using pseudonymous keys in the coded data, omitting local paths and secret values, identifying cases in this paper only by issue number, and paraphrasing identifying quotes. The full downloaded payloads and the internal source map are kept private. A second risk is dual use: describing how control fails could inform misuse. We judge the benefit higher, because these failures are already public, the paper describes no new exploit, and its purpose is to help designers close the gaps it names. The study involved no recruitment, no interaction with people, and no private data.

\section*{Data Availability}

The coded episode file, screening ledger, codebook, audit log, and analysis scripts are available from the author on request. Downloaded issue payloads are withheld because they reproduce account handles, local paths, and possibly sensitive strings.

\bibliographystyle{ACM-Reference-Format}
\bibliography{references}

\appendix

\section{Episode Index}
\label{app:episodes}

Table~\ref{tab:episodes} lists the 30 episodes with their source issue, failure stage, gap type, and strongest mechanism evidence. Episodes that share a family identifier arise from the same technical defect.

\begin{table*}[t]
  \caption{The 30 episodes. Source is the GitHub issue (repository and number); ``(c)'' marks an episode reported in a comment on that issue. Gap types: UE unclear enforcement, UC unseen capability, SH soft instruction treated as hard, BH broken hard control. Evidence: R reporter's account only, P reproduced, C confirmed in code or pull request, M confirmed by maintainer.}
  \label{tab:episodes}
  \small
  \begin{tabular}{@{}lllllll@{}}
    \toprule
    ID & Product & Source & Stage & Gap & Evidence & Family \\
    \midrule
    E001 & Claude Code & claude-code \#74547 & saved state & UE & R & F001 \\
    E002 & Claude Code & claude-code \#81138 & interface & UC & P & F002 \\
    E003 & Claude Code & claude-code \#85294 & saved state & UE & P & F002 \\
    E004 & Claude Code & claude-code \#85548 & interface & UE & P & F003 \\
    E005 & Claude Code & claude-code \#87971 & instruction & UE & P & F004 \\
    E006 & Claude Code & claude-code \#87971 (c) & executor & UE & R & F004 \\
    E007 & Claude Code & claude-code \#87971 (c) & instruction & SH & R & F004 \\
    E008 & Claude Code & claude-code \#90450 (c) & instruction & UE & P & F004 \\
    E009 & Claude Code & claude-code \#92911 & permission & UE & P & F005 \\
    E010 & Codex CLI & codex \#22124 & saved state & BH & P & F006 \\
    E011 & Codex app-server & codex \#35894 & interface & UE & C & F007 \\
    E012 & Codex Desktop & codex \#42902 & interface & UC & P & F008 \\
    E013 & Codex Desktop & codex \#43062 & recovery & UE & R & F009 \\
    E014 & Codex Desktop & codex \#43840 & saved state & BH & R & F010 \\
    E015 & Codex Desktop & codex \#43885 & recovery & UE & R & F011 \\
    E016 & Claude Code & claude-code \#60705 & instruction & SH & R & F012 \\
    E017 & Gemini CLI & gemini-cli \#20355 & permission & BH & P & F013 \\
    E018 & Gemini CLI & gemini-cli \#23411 & executor & UE & R & F014 \\
    E019 & Claude Code & claude-code \#87348 & saved state & UC & P & F015 \\
    E020 & Gemini CLI & gemini-cli \#26856 & executor & SH & R & F016 \\
    E021 & Codex Desktop & codex \#34331 & permission & UE & P & F017 \\
    E022 & Codex Desktop & codex \#38886 & permission & UE & P & F017 \\
    E023 & Codex Desktop & codex \#42523 & recovery & UE & P & F018 \\
    E024 & Codex Desktop & codex \#42027 & recovery & UE & C & F019 \\
    E025 & Codex Desktop & codex \#43142 & recovery & UE & M & F019 \\
    E026 & Claude Code & claude-code \#83775 & executor & SH & R & F020 \\
    E027 & Claude Code & claude-code \#74651 & interface & UC & P & F021 \\
    E028 & Claude Code & claude-code \#63356 & executor & SH & R & F022 \\
    E029 & Gemini CLI & gemini-cli \#25863 & executor & UE & R & F023 \\
    E030 & Codex Desktop & codex \#43675 & permission & UC & C & F024 \\
    \bottomrule
  \end{tabular}
\end{table*}

\end{document}